# Influence of Diode Laser Intensity Modulation on Photoacoustic Image Quality for Oral Soft Tissue Imaging

**Rini Widyaningrum[1], Rellyca Sola Gracea[2], Dewi Agustina[3], Munakhir Mudjosemedi[2], Mitrayana[4]**

[1] Doctoral Program, Faculty of Dentistry, Universitas Gadjah Mada, Yogyakarta, Indonesia.
[2] Department of Dentomaxillofacial Radiology, Faculty of Dentistry, Universitas Gadjah Mada, Yogyakarta, Indonesia.
[3] Department of Oral Medicine, Faculty of Dentistry, Universitas Gadjah Mada, Yogyakarta, Indonesia.
[4] Department of Physics, Faculty of Mathematics and Natural Sciences, Universitas Gadjah Mada, Yogyakarta, Indonesia.

**ABSTRACT**

Imaging technologies have been developed to assist physicians and dentist in the detection of various diseases. Photoacoustic imaging (PAI) is a new imaging technique that shows great promise to image soft tissues. The prototype of PAI system in this study utilized a non-ionizing 532 nm continue-wave diode laser illumination to image oral soft tissue. The aim of this study was to investigate the effect of diode laser intensity modulation to the Photoacoustic (PA)-image quality. Samples in this study were oral soft tissues from six Sprague Dawley rats. This samples were placed in wax and then imaged by using the PAI system. To determine the optimum duty cycle of laser intensity modulation, the laser exposure for oral soft tissue imaging was set in various duty cycles, i.e. 16%, 24%, 31%, 39%, and 47 %. The Kruskal-Wallis test followed with Mann–Whitney post hoc analysis revealed there was statisticallly significant differences ($p<0.05$) between PA-images quality produced by using 16-47% duty cycle of laser intensity modulation. The PAI system built in this study was able to image oral soft tissue. The optimum duty cycle of laser intensity modulation used in the PAI system for oral soft tissue imaging was 39%.

*Correspondence:* Rini Widyaningrum, Doctoral Program, Faculty of Dentistry, Universitas Gadjah Mada, Jl. Denta, Sekip Utara, Sleman, Yogyakarta, Indonesia. Email: rinihapsara@ugm.ac.id, Phone number: +6281578711722

## INTRODUCTION

Radiography examination and imaging technology are integral parts in dental and medical treatment[1]. The imaging technology is advancing from X ray radiography that utilized the ionizing radiation into several imaging technology that utilized non-ionizing radiation, i.e. ultrasonography (USG), magnetic resonance imaging (MRI), and photoacoustic imaging (PAI). X-ray radiography is most suitable to image hard tissue, but this imaging technique has limitation to image soft tissue. Meanwhile the photoacoustic imaging combines the merits of both optical and acoustic imaging, and become a promising structural, functional and molecular imaging modality for a wide range of biomedical applications[2–4], especially to overcome the limitation of previous imaging modality in soft tissues imaging.

Photoacoustic imaging (PAI) is a new emerging imaging technique based on the photoacoustic effect. The photoacoustic effect was first observed by Alexander G. Bell in 1880[2,5,6], when he found that absorption of electromagnetic (EM) waveforms, such as radio-frequency (rf) and optical waves can generate transient acoustic signals in media. Such absorption leads to local heating in materials called thermoelastic expansion, which can produce (photo-) acoustic waves[2].

Different from X-ray radiography, the PAI utilized a non-ionizing electromagnetic waves that commonly generated by laser or other optical energy source[7]. This imaging technique capable to image organs, tissues and cellular structures with good resolution and excellent contrast[2–4]. Photoacoustic techniques were initially studied in non-biological fields such as physics and chemistry. Since Theodore Bowen introduced PAI technique as a biomedical imaging modality in 1981, this technology has been developing quite rapidly[4]. Nowadays in Indonesia, studies on photoacoustic effect has been applied in the field of physics[8], biology[9], agriculture[10], and biomedical field[11,12]. Whilst application of photoacoustic effect as an imaging technology has never been studied in Indonesia yet[13].

Since different biological tissues have different absorption coefficient, we can rebuild the distribution of optical energy deposition and ultimately obtain images of the biological tissues by measuring the acoustic signals with an ultrasonic transducers[2] or other acoustic detector such as microphone. Biological tissues contain several kinds of endogenous chromophores such as hemoglobin, melanin, and lipids. These endogenous chromophores can absorb the electromagnetic energy from laser or other electromagnetic source, and then generate laser-induced acoustic that so-called photoacoustic (PA) waves[4,14]. The endogenous chromophores have stronger absorption coefficients in comparison with other tissue constituents[4], so that they may act as an endogenous contrast agent in photoacoustic imaging.

The PAI system is commonly built by integrating ultrasound transducer as detector with a pulsed laser or a continuous-wave (CW) laser with intensity modulation as EM source[15]. In this study, we have designed and demonstrated a low cost photoacoustic imaging system (reflection mode) in which a 532nm continue-wave low energy diode laser with maximum output power 200mW used as EM source. To significantly reduce the cost, the diode laser was combined with condenser microphone and a customized XY stage in the system. Therefore to determine the optimum laser modulation to produce the best quality of photoacoustic (PA-) image in this study, they were taken by using various laser duty cycles, i.e. 16%, 24%, 31%, 39%, and 47%. In addition to get the best PA-image quality, the optimum duty cycle applied in the system is also expected will not give any bio-effect on oral soft tissue samples in this study.

## MATERIALS AND METHODS

Ethical approval for this study was received from the Medical and Health Research Ethics Committee, Faculty of Medicine, Universitas Gadjah Mada-Dr Sardjito General Hospital, Yogyakarta, Indonesia (Ref: KE/FK/0285/EC/2017).

The photoacoustic (PAI) system built in this study is shown in Fig. 1. The EM source in the system was a 532nm diode laser (maximum output power 200mW). A convex lens with focal length of 100mm was used to focus the laser beam onto the imaged area on the tissue surface. The distance between the lens and the imaged sample was adjusted to achieve the optimal focusable laser beam. The PAI system used a condenser microphone ECM8000 (Behringer, Germany) as photoacoustic signal detector, connected with a portable computer by using an USB audio interface with mic preamplifier. A customized XY stage was built to move the sample in the two-dimensional (2-D) X and Y direction to obtain PA-images. A portable computer was used to control the system, process data acquisition, perform the PA-image reconstruction, and display the PA-images.

Sample of this study was oral soft tissue placed in wax. The oral soft tissue was a sagittal slice of Sprague Dawley rats tongue. Photoacoustic imaging was performed ex vivo by using the PAI system without any exogenous contrast agents.

**Laser modulation**

To generate photoacoustic signals from samples, the PAI system used intensity-modulated CW diode laser. The laser modulation applied the Pulse Width Modulation (PWM) technique. The laser light was turned on and off periodically to form a square wave fluctuation in the laser transistor-transistor logic (TTL) input, by set the duty cycle of laser irradiation on a single frequency of 17 kHz. A microcontroller connected with a portable computer was used to set the laser intensity-modulation, as shown in Fig. 1.

**Imaging of the Sample**

The PA-imaging was done by placing sample on XY stage table and then illuminated by using intensity-modulated diode laser in the PAI system. The surface of the sample was sagittal plane of oral soft tissue. It was arranged evenly flat, parallel with horizontal plane. The PA-imaging process was done in the Region of Interest (ROI) on the sample. The ROI marked with white-square in the sample, as shown in Fig. 2A. To ensure that the PAI-images produced by the system was able to depict the biological tissue from non-biological material (wax) around it, the ROI in this study includes both of tissue area and wax area.

During PA-imaging, the XY stage moved the sample in 200 μm distance for each step, and the system was set to record the PA-signal in 1 second duration from each point in the ROI. Data acquisition employ a single-frequency detection by using a condenser microphone as detector in this system. The detector recorded the PA-signals and then converted them into sound spectrum by employing the fast fourier transform (FFT). The peak of acquired PA-signal intensity around frequency of 17 kHz was taken to form the element of single point in the PA-image. PA-image reconstruction was based on the PA signal intensity recorded from each point in the ROI.

**The PA-Image Quality**

The PA-image quality in this study was assessed by 2 observers based on simple criteria set in this study (Table 1). The observers were calibrated before the PA-image quality assessment. The Kappa coefficient was calculated to determine inter-examiner agreement regarding the scores of PA-image quality.

**Data Analysis**

To determine the optimum laser modulation used in the PAI system to image oral soft tissue, PA-images was produced from each sample by using five gradual duty cycle of laser intensity-modulation, i.e. 16%, 24%, 31%, 39%, and 47 %. Therefore there were five PA-images produced from each sample in this study, and total of thirty PA-images were collected from six samples in this study. Photoacoustic image quality data was then separated into 5 groups based on the duty cycle of laser intensity modulation used to produce them. The image quality between PA-images group then were compared by using The Kruskal-Wallis test followed with Mann–Whitney post hoc analysis. The significance level was set at 5% ($p<0.05$).

# RESULTS

Figure 2 shows a sample and PA-images taken from ROI of the same sample by using various gradual duty cycle of laser intensity modulation. The figure shows that various gradual duty cycle applied in laser intensity modulation were able to produce PA-images with various quality.

Bright color (yellow and red) represent higher PA-signal intensity than the dark (blue) ones. The oral soft tissue is seen as a white object surrounded by green media in the sample (Fig. 2A), and it can be seen as yellow area (Fig. 2C), or yellow and red area (Fig. 2D, 2E, and 2F) in PA-images. As seen in Fig. 2, the biological tissue which was oral soft tissue produced yellow and red area in PA-image, and the surrounding non biological material produced blue area. So, the biological tissues have higher PA-intensity than the non-biological material. The sample cannot clearly visualized in PA-images produced by using the lowest (16%) duty cycle of laser intensity modulation (Fig. 2B). Duty cycle 24-47% may produce visible soft tissue image, since it is show as yellow and red area in Fig. 2D, 2E, and 2F. These PA-images shows same shape with the real object in the ROI at Fig. 2A.

PA-intensity tends to increase along with the increasing of laser intensity-modulation. The higher duty cycle of laser modulation, the higher PA-intensity generated from the sample. Therefor PA-image produced by using the highest duty cycle of laser intensity modulation shows widest red area in PA-image (Fig. 2F) compared with previous PA-images produced by using lower duty cycle (Fig 2B, 2C, 2D, and 2E).

It can be seen from Fig. 2 that duty cycle 39% was the most appropriate to be applied for laser intensity modulation, as it produced more visible image of oral soft tissue (Fig. 2E) compared with other PA-images (Fig. 2B, 2C, 2D, and 2F) that produced by using lower or higher duty cycle. The outline of oral soft tissue in Fig. 2E is more well-defined, therefore it can be distinguished from the surrounding non-biological material.

Furthermore, highest duty cycle of laser intensity modulation in this study may produced unwanted bio-effect that is burning effect. This effect was found both in oral soft tissue as well as non-biological material that used as sample. PA-image from burnt samples show low image quality with red-artefact so that it may not visualize the real object (oral soft tissue), as it can be seen in PA-images produced by using the highest (47%) duty cycle of laser intensity modulation (Fig. 2F).

According to all (30) of PA-images quality assessment, the Kappa value of inter-observer reliability was 0,606 (p<0.001), which indicates good agreement among the observers. Table 2 shows that PA-image quality is varied among groups. The Kruskal-Wallis test followed with Mann–Whitney post hoc analysis was applied to evaluate the differences of PA-image quality between groups. The statistical analysis (Table 2) revealed that there were significant differences (p<0.05) in PA-images quality among groups, especially between group of duty cycle 16% vs 31%, 16% vs 39%, and 16% vs 47%.

**DISCUSSION**

Fundamentally, the photoacoustic technique measures the conversion of electromagnetic energy into acoustic pressure waves. In biomedical PAI, the tissue is irradiated with a pulsed laser, resulting in the generation of an ultrasound wave as a consequence of the optical absorption followed by rapid thermal (or thermoelastic) expansion and subsequent relaxation of tissue[4,16]. The thermal expansion need to be time variant so that tissue or material irradiated by electromagnetic wave can generate acoustic wave. This condition can be reached by using either a pulsed laser or a continuous-wave (CW) laser with intensity modulation as EM source in the PAI system. The laser intensity modulation for CW-wave lasers can be applied at a constant or variable frequency[15].

Pulsed laser is the most widely used in PAI system as it produces higher signal to noise ratio rather than the CW laser. However, these lasers are large size, bulky, lack of practical utility, and thus not easily portable. In addition, these lasers are expensive, require external cooling systems, need for re-alignment as well as to be operated by technically experienced personnel. That facts are undoubtedly inhibits their practical use[17]. On the contrary, CW diode laser offer attractive alternatives as an excitation source of PAI system as it is smaller, inexpensive, compact, and durable [14]. Recently, the feasibility of diode laser-based PAI has been explored in vivo for superficial imaging applications of blood vessels and skin [3,14,18]. Furthermore, this study also shows the feasibility of the PAI system based on diode laser intensity modulation to image oral soft tissue ex vivo.

The biomedical photoacoustic imaging is commonly used ultrasound transducer as detector to record the PA-signal generated by the sample. The PAI system built in this study used audio-sonic condenser microphone that generally used as detector in gas-microphone photoacoustic technique[19,20]. This detector is simple and low cost. Yet the detector need to be develop further for in-vivo experiment.

Instead of using ionizing radiation that obviously utilized in X-ray radiography, PAI utilized non-ionizing laser irradiation as EM source[15]. Since the PAI technique apply the principle of conversion light into sound (or acoustic) wave, the PA-image is reconstructed through an imaging process from optically induced acoustic (or ultrasound) signals generated from the sample. Every single point in the PAI-image in this study is represent the PA-signal intensity detected from the same point in the ROI. Higher PA-intensity show brighter colour in PA-image. The PA-images shows gradual colour from dark blue which represent the lowest PA-signal intensity, then it turns into brighter colours which are yellow and red which represent higher PA-signal intensity. The higher duty cycles employed to modulated laser in PAI system, the higher intensities of PA-signal generated from samples (Fig. 2), therefor PA-image that was reconstructed from them showed brighter color as well.

This study shows that laser modulation may influence the PA-image quality. Therefor the laser modulation has to be applied in optimum duty cycle to get the best PA-Image quality. Result of this study (Table 2) reveals that there is a significant difference ($p<0.05$) in PA-Image quality produced by using duty cycle 16-47%, especially between group of 16% with 31%, 39%, and 47% group. Low intensity (duty cycle 16-24%) of 532nm-diode laser illumination in this study was not sufficient to generate PA-signals to be produced as a PA-Image of the sample. On the contrary, very high intensity of laser illumination may produces unwanted bio-effects[21] e.g. burnt effect, that may degrade the PA-image quality, especially when the laser is focused onto a very small spot[21]. Besides that, the PA image quality may be also affected by noise during data acquisition. Generally, images may be degraded by some form of impulse noise that mostly caused during the process of image acquisition, transmission through communication media and storage in the physical devices [22]. In this study, noise that degrade the PA-images may be received during data acquisition and produced by burning effect caused by high intensity laser modulation.

Further research should be done by considering reduction of noise both in image acquisition and image reconstruction. Development in both of hardware and software component as well as sample treatment are needed to prevent the unwanted biological effect of laser irradiation to the sample and to reduce noises that may produces artefact in PA-images.

## CONCLUSION

In conclusion, the effect of laser intensity-modulation was statistically significant to the PA-image quality, especially between PA-image produced by the lowest duty cycle (16%) with higher duty cycle (31%, 39%, and 47%) of laser intensity modulation. The optimum duty cycle to modulated 532nm diode laser for oral soft tissue imaging in this study was 39%. Yet much effort must still be invested to develop the system and imaging method in this study.

## ACKNOWLEDGEMENT

This study was supported by Schollarship for Graduate Program from The Indonesian Ministry of Research, Technology, and Higher Education and also supported by research fund from The Faculty of Dentistry Universitas Gadjah Mada. The authors declare no potential conflict of interests regarding the authorship and/or publication of this article. We wish to thank Eddy Kurniawan and Harry Miyosi Silalahi for their assistance, especially with Fig. 1.

**FIGURE**

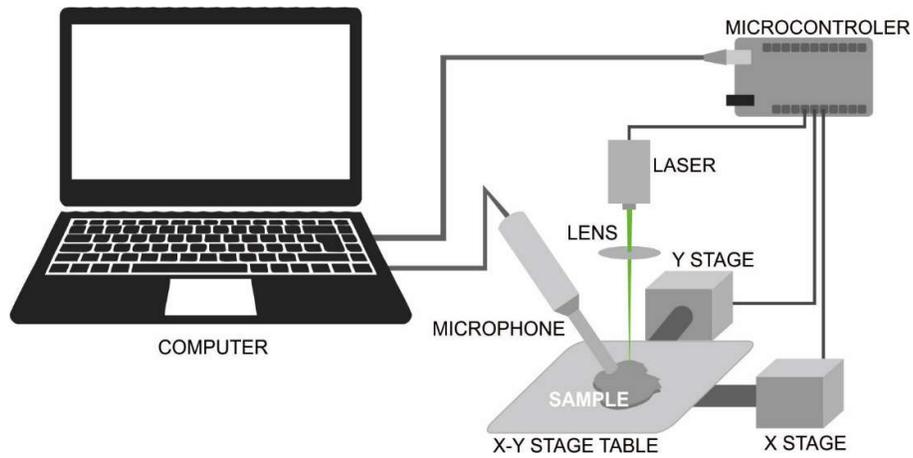

Figure 1. Schematic of the PAI system. The system used a 532nm diode laser as EM source, a condenser microphone as detector, and a customized XY stage to move the sample during PA-imaging.

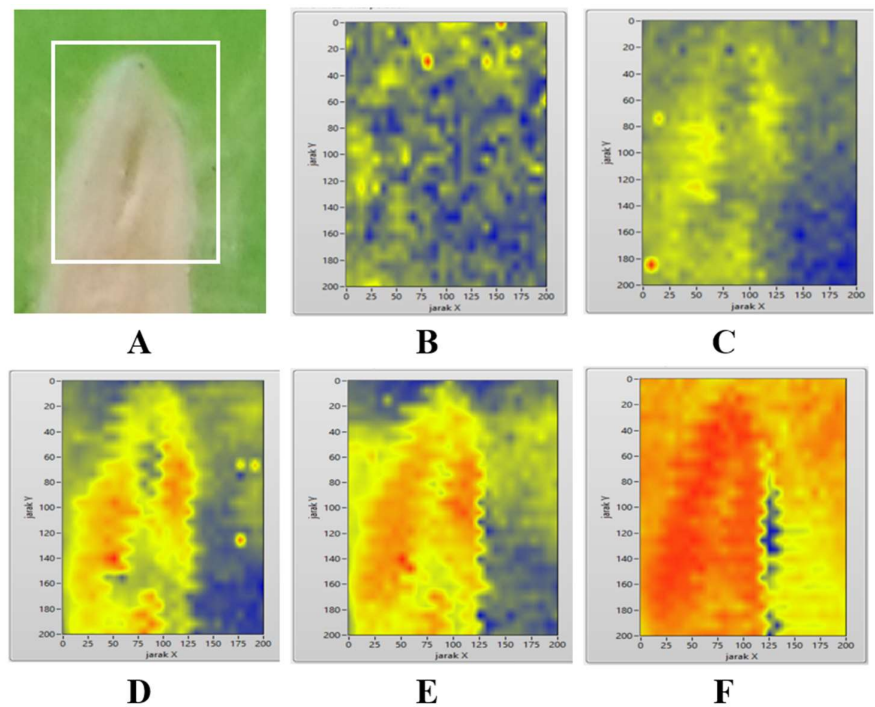

Figure 2. Sample (A) and PA-images produced with various duty cycle of laser-intensity modulation: 16% (B), 24% (C), 31% (D), 39% (E), and 47% (F). The ROI is marked with white-square in the sample (A).

**TABLE**

Table 1. Assessment of PA-Image Quality

| Score | Criteria |
|---|---|
| 1 | The oral soft tissue is clearly visualized in the PA-image and can be differed clearly from wax media around it. The outline of oral soft tissue is well-defined. |
| 2 | The oral soft tissue is visualized in the PA-image, but cannot be differed clearly from wax media around it. Artefact may found in the PA-image. |
| 3 | The oral soft tissue is not visualized in the PA-image. Artefact may obscures the oral soft tissue image. |

Table 2. Statistical Analysis

| | | n | PA-Image Quality Me (Min – Max) | P Value |
|---|---|---|---|---|
| **Duty Cycle of Laser Intensity Modulation** | 16 % | 6 | 1 (1-1) | $P < 0.05$ |
| | 24 % | 6 | 1 (1-3) | |
| | 31 % | 6 | 2 (2-3) | |
| | 39 % | 6 | 2.5 (2-3) | |
| | 47 % | 6 | 2 (2-3) | |

Kruskal-Wallis test. Mann-Whitney post hoc: 16% vs 24% p=0.394; 16% vs 31% p<0.05; 16% vs 39% p<0.05; 16% vs 47% p<0.05; 24% vs 31% p=0.132; 24% vs 39% p=0.065; 24% vs 47% p=0.093; 31% vs 39% p=0.394; 31% vs 47% p=0.699; 39% vs 47% p=0.699